\documentclass[twocolumn,english,superscriptaddress,showpacs,prl]{revtex4}
\usepackage[utf8]{inputenc}
\usepackage{color}
\usepackage{babel}
\usepackage{bm}
\usepackage{amsmath}
\usepackage{amsthm}
\usepackage{amssymb}
\usepackage{graphicx}
\usepackage[colorlinks=true,citecolor=blue]{hyperref}

\usepackage{pdfpages}
\usepackage{times}

\newtheorem{thm}{Theorem}
\newtheorem{lem}{Lemma}

\begin{document}

\title{Entanglement area laws for long-range interacting systems}

\author{Zhe-Xuan Gong}
\email{gzx@umd.edu}

\selectlanguage{english}%

\affiliation{Joint Quantum Institute, NIST/University of Maryland, College Park, Maryland 20742, USA}

\affiliation{Joint Center for Quantum Information and Computer Science, NIST/University of Maryland, College Park, Maryland 20742, USA}

\author{Michael Foss-Feig}

\affiliation{United States Army Research Laboratory, Adelphi, MD 20783, USA }

\author{Fernando G. S. L. Brandão}

\affiliation{IQIM, California Institute of Technology, Pasadena CA 91125, USA}

\author{Alexey V. Gorshkov}

\affiliation{Joint Quantum Institute, NIST/University of Maryland, College Park, Maryland 20742, USA}

\affiliation{Joint Center for Quantum Information and Computer Science, NIST/University of Maryland, College Park, Maryland 20742, USA}
\begin{abstract}
We prove that the entanglement entropy of any state evolved under an arbitrary $1/r^{\alpha}$ long-range-interacting $D$-dimensional lattice spin Hamiltonian cannot change faster than a rate proportional to the boundary area for any $\alpha>D+1$. We also prove that for any $\alpha>2D+2$, the ground state of such a Hamiltonian satisfies the entanglement area law if it can be transformed along a gapped adiabatic path into a ground state known to satisfy the area law. These results significantly generalize their existing counterparts for short-range interacting systems, and are useful for identifying dynamical phase transitions and quantum phase transitions in the presence of long-range interactions. 
\end{abstract}


\maketitle
Quantum many-body systems often have approximately local interactions, and this locality has profound effects on the entanglement properties of both ground states and the states created dynamically after a quantum quench. For example, the entanglement entropy, defined as the entropy of the reduced state of a subregion, often scales as the boundary area of the subregion for ground states of short-range interacting Hamiltonians \cite{eisert_colloquium:_2010}. This ``area law'' of entanglement entropy is in sharp contrast to the behavior of thermodynamic entropy, which typically scales as the volume of the system. While the study of area laws originates from black hole physics \cite{bekenstein_black_1973,hawking_black_1974}, area laws have received considerable attention recently in the fields of quantum information and condensed matter physics. In particular, area laws are known to be closely related to the velocity of information propagation in quantum lattices \cite{lieb_finite_1972}, quantum critical phenomena and conformal field theory \cite{calabrese_entanglement_2004}, the efficiency of classical simulation of quantum systems \cite{vidal_efficient_2003}, topological order \cite{kitaev_topological_2006}, and many-body localization \cite{nandkishore_many-body_2015}.

However, the description of many-body systems in terms of local interactions is often only an approximation, and not always a good one; in numerous systems of current interest, ranging from frustrated magnets and spin glasses \cite{ruderman_indirect_1954,binder_spin_1986} to atomic, molecular, and optical systems \cite{saffman_quantum_2010,yan_observation_2013,aikawa_bose-einstein_2012,islam_emergence_2013,britton_engineered_2012,douglas_quantum_2015}, long-range interactions are ubiquitous and lead to qualitatively new physics, e.g.\ giving rise to novel quantum phases and dynamical behaviors \cite{kastner_diverging_2011,eisert_breakdown_2013,yao_realizing_2013,yao_many-body_2014,vodola_kitaev_2014,gong_kaleidoscope_2016,maghrebi_continuous_2015,neyenhuis_observation_2016}, and enabling speedups in quantum information processing \cite{avellino_quantum_2006,richerme_non-local_2014,jurcevic_quasiparticle_2014,eldredge_fast_2016,foss-feig_entanglement_2016}. Particles in these systems generally experience interactions that decay algebraically ($\sim1/r^{\alpha}$) in the distance ($r$) between them. As might be expected, $\alpha$ controls the extent to which the system respects notions of locality developed for short-range interacting systems: For $\alpha$ sufficiently small, it is well established \cite{eisert_breakdown_2013} that locality may be completely lost, and for $\alpha$ sufficiently large there is ample numerical and analytical evidence \cite{koffel_entanglement_2012,nezhadhaghighi_quantum_2013,vodola_long-range_2015,gong_topological_2016} that area laws may persist. However, there is currently no general and rigorous understanding of when area laws do or do not survive the presence of long-range interactions.

The modern understanding of area laws draws heavily from several rigorous proofs, all of which require some restrictions on the general setting discussed above. As the most notable example, Hastings \cite{hastings_area_2007} proved that ground states of one-dimensional (1D) gapped Hamiltonians with finite-range interactions satisfy the area law. A subsequent development was made later in Refs.\,\cite{brandao_area_2013,brandao_exponential_2014}, which proved that states in 1D with exponentially decaying correlations between any two regions (a set that includes the ground states of gapped short-range interacting Hamiltonians) must satisfy the area law. Generalizing these proofs to include long-range interacting Hamiltonians is, however, rather difficult. For example, it is a well-known challenge to generalize Hastings' proof of the area law \cite{hastings_area_2007} to higher dimensions \cite{arad_improved_2012}, and long-range interacting systems are in some sense similar to higher-dimensional short-range interacting systems. In addition, since ground states of gapped long-range interacting systems can have power-law decaying correlations \cite{hastings_spectral_2006,schachenmayer_dynamical_2010,maghrebi_causality_2016}, one would need to relax the condition of exponentially decaying correlations in the proof of Refs.\,\cite{brandao_area_2013,brandao_exponential_2014} to algebraically decaying correlations. However, this relaxation invalidates the proof, as there exist 1D states with sub-exponentially decaying correlations that violate the area law \cite{hastings_random_2015}.

To circumvent these challenges in proving area laws for long-range interacting systems, here we employ a ``dynamical'' approach. Specifically, we prove that a state satisfies the area law if it can be dynamically created in a finite time by evolving a state that initially satisfies the area law under a long-range interacting Hamiltonians \cite{van_acoleyen_entanglement_2013}. We then use the powerful formalism of quasi-adiabatic continuation \cite{hastings_quasiadiabatic_2005} to relate such a state to the ground state of a spectrally gapped long-range interacting Hamiltonian. This strategy is made possible by the recent proof of Kitaev's small incremental entangling (SIE) conjecture \cite{bravyi_upper_2007,van_acoleyen_entanglement_2013}, and by significant recent improvements in Lieb-Robinson bounds \cite{lieb_finite_1972} for long-range interacting systems \cite{foss-feig_nearly_2015,gong_persistence_2014}.

The manuscript is divided into two proofs of two different area laws, the latter of which builds on the former. The first area law states that for any initial state, the entanglement entropy of a subsystem cannot change faster than a rate proportional to the subsystem's area. This statement is known to hold for short-range interacting systems \cite{van_acoleyen_entanglement_2013,ho_entanglement_2015}, and we have generalized it to systems with interactions decaying faster than $1/r^{D+1}$. A direct implication of this new area law is that matrix-product-state calculations of quench dynamics are expected to remain efficient at relatively short times for most $1/r^{\alpha}$ Hamiltonians with $\alpha>D+1$. Moreover, the proof of our area law also suggests that for $\alpha\le D$, it might be possible for the entanglement entropy to change from an area law to a volume law in a finite time, thus indicating the onset of a dynamical phase transition \cite{heyl_dynamical_2013}.

Our second area law states that if a Hamiltonian has interactions decaying faster than $1/r^{2D+2}$, then its ground state satisfies the area law if it can be connected to an area-law state by adiabatically deforming the Hamiltonian. Here adiabaticity implies a finite energy gap during the adiabatic evolution and requires interactions to still decay faster than $1/r^{2D+2}$. This area law leads to two new insights: (1) The entanglement area law for the ground state of a gapped short-range interacting Hamiltonian will remain stable if we add long-range interactions without closing the gap. For certain frustration-free Hamiltonians, including Kitaev's toric code \cite{kitaev_alexei_anyons_2006} and the Levin-Wen model \cite{levin_string-net_2005-1}, the area law is strictly implied for $\alpha>2D+2$ due a proven stability of the gap for interactions decaying faster than $1/r^{D+2}$ \cite{michalakis_stability_2013}. Thus the short-range nature of interactions, believed to be crucial for area laws, is in fact not necessary. (2) The entanglement area law might be violated without destroying the energy gap or making the energy non-extensive by using $1/r^{\alpha}$ interactions with $D<\alpha<2D+2$ . Thus there may exist exotic quantum phase transitions between gapped phases, challenging the conventional wisdom that quantum phase transitions cannot take place between gapped phases in an adiabatic evolution \cite{chen_local_2010-1}.

\emph{Main results}.— In this manuscript, we consider the following Hamiltonian $H$ on a D-dimensional finite or infinite lattice 
\begin{equation}
H=\sum_{i,j}h_{ij},\hspace{1em}\left\Vert h_{ij}\right\Vert \le1/r_{ij}^{\alpha}\quad(i\ne j).\label{eq:H}
\end{equation}
Here, $h_{ij}$ is an operator acting on sites $i$ and $j$ that can be time-dependent, $\lVert h_{ij}\rVert$ denotes the operator norm (largest-magnitude of an eigenvalue) of $h_{ij}$, and $r_{ij}$ represents the distance between sites $i$ and $j$. The maximum Hilbert space dimension for any site is assumed to be finite and denoted by $d$. The strength of the on-site interaction $h_{ii}$ can be arbitrary, and is unimportant in the following area laws and proofs.

We define the entanglement entropy of a state $|\psi\rangle$ with respect to a subregion $V$ by $S_{V}(|\psi\rangle)\equiv-\text{tr}[\rho_{V}\log\rho_{V}]$, where $\rho_{V}=\text{tr}_{\bar{V}}(|\psi\rangle\langle\psi|)$ and $\bar{V}$ is the complement of $V$. We will use $\partial V$ to denote the set of sites at the boundary of $V$, and $|V|$ to denote the number of sites in the set $V$. To clarify the presentation without sacrificing rigor, we will frequently use the identification $g(x)=\mathcal{O}(x)$ if there exists finite positive constants $c$ and $x_{0}$ such that $g(x)\le cx$ for all $x\ge x_{0}$. The constants $c$ and $x_{0}$ may be different each time the $\mathcal{O}$-notation appears, but will not depend on anything other than the lattice geometry and fixed parameters $\alpha$, $D$, $d$, and $\Delta$ (introduced later). We now state our first area law: 
\begin{thm}
(Area law for dynamics) For any state $|\psi\rangle$ under the time evolution of $H$ defined in Eq.\,\eqref{eq:H} with $\alpha>D+1$, 
\begin{equation}
\left|\frac{dS_{V}(|\psi(t)\rangle)}{dt}\right|\le\mathcal{O}(|\partial V|).
\end{equation}
\end{thm}
\noindent To prove Theorem 1, let us introduce the following lemma, which can be directly obtained from the Kitaev's SIE conjecture recently proven in Ref.\,\cite{van_acoleyen_entanglement_2013}. 
\begin{lem}
If $H=\sum_{Z}h_{Z}$ with $h_{Z}$ acting on a set of sites $Z$, then for any state $|\psi\rangle$ 
\begin{equation}
\left|\frac{dS_{V}(|\psi(t)\rangle)}{dt}\right|\le18\log(d)\!\!\!\!\sum_{Z,\thinspace Z\cap V\ne\emptyset\thinspace\&\thinspace Z\cap\bar{V}\ne\emptyset}\left\Vert h_{Z}\right\Vert |Z|.\label{eq:SIE}
\end{equation}
\end{lem}
\noindent Roughly speaking, this lemma tells us that the entanglement entropy at most changes at a rate proportional to the total strength of interactions that cross the boundary of $V$.

With the help of Lemma 1, the proof of Theorem 1 reduces to the proof of $\sum_{i\in V,j\notin V}\left\Vert h_{ij}\right\Vert \le\mathcal{O}(|\partial V|)$. Let us now assign a coordinate $(\bm{x}_{i},r_{i})$ to each site $i\in V$, with $\bm{x}_{i}$ measuring the directions parallel to the boundary, and $r_{i}$ measuring the distance of $i$ to the boundary (rounded down to the next integer). Upon bounding the sum by a D-dimensional integral, it is straightforward to show that for a given $i\in V$, $\sum_{j\notin V}\left\Vert h_{ij}\right\Vert \le\mathcal{O}(r_{i}^{D-\alpha})$ . Since for a given value of $r_{i}$, the possible choices of $\bm{x}_{i}$ is at most proportional to $|\partial V|$, it follows that $\sum_{i\in V,j\notin V}\left\Vert h_{ij}\right\Vert \le\mathcal{O}(|\partial V|)\sum_{r=1}^{\infty}r^{D-\alpha}$. Theorem 1 is then proven because $\sum_{r=1}^{\infty}r^{D-\alpha}$ converges for $\alpha>D+1$. Note that the method used here is an improvement over a similar method used in Ref.\,\cite{van_acoleyen_entanglement_2013}, which if used will lead to the condition $\alpha>D+2$ instead.

To connect from this dynamical area law to a ground-state area law, we now introduce the formalism of quasi-adiabatic continuation. Assume that there is a continuous family of Hamiltonians 
\begin{equation}
H(s)=(1-s)H(0)+sH(1),\label{eq:Hs}
\end{equation}
parameterized by $s\in[0,1]$ with each $H(s)$ being a time-independent Hamiltonian satisfying Eq.\,\eqref{eq:H} and having a unique ground state $|\psi_{0}(s)\rangle$ and a finite energy gap of at least $\Delta$. As shown in Ref.\,\cite{hastings_quasiadiabatic_2005}, the evolution (or continuation) of $|\psi_{0}(s)\rangle$ from $s=0$ to $s=1$ is governed by an effective Hamiltonian $\mathcal{D}(s)$, given by the ``Schrodinger equation'' $d|\psi_{0}(s)\rangle/ds=-i\mathcal{D}(s)|\psi_{0}(s)\rangle$. We emphasize that the evolution of $|\psi_{0}(s)\rangle$ is not governed by $H(s)$, because despite the existence of a finite gap $\Delta$, to adiabatically evolve under $H(s)$ from $|\psi_{0}(0)\rangle$ to $|\psi_{0}(1)\rangle$ \emph{exactly} requires an infinite evolution time, in contrast to the unity time needed for the evolution under $\mathcal{D}(s)$. As a result, the evolution of $|\psi_{0}(s)\rangle$ under $\mathcal{D}(s)$ is usually called quasi-adiabatic continuation \cite{hastings_quasiadiabatic_2005}.

For a given $H(s)$, the choice of $\mathcal{D}(s)$ is not unique, and here we choose a convenient form given in Ref.\,\cite{bravyi_topological_2010},

\begin{equation}
\mathcal{D}(s)=-i\int_{-\infty}^{\infty}f(\Delta t)e^{iH(s)t}\frac{\partial H(s)}{\partial s}e^{-iH(s)t}dt.\label{eq:qa}
\end{equation}
Here, $f(x)$ belongs to a family of \emph{sub-exponentially} decaying functions, meaning that for any $\delta<1$, there exists an $x$-independent constant $c_{\delta}$ such that $|f(x)|\le c_{\delta}\exp(-|x|^{\delta})$ {[}the explicit form of $f(x)$ is not important{]}. The $\mathcal{D}(s)$ given in Eq.\,\eqref{eq:qa} has a remarkable feature: if $H(s)$ is a short-range interacting Hamiltonian {[}Eq.\,\eqref{eq:Hs} in the $\alpha\rightarrow\infty$ limit{]}, then $\mathcal{D}(s)$ contains interactions that decay sub-exponentially with distance, approximately inheriting the locality of the underlying interactions \cite{hastings_quasiadiabatic_2005}. For a finite but suitably large $\alpha$, it is reasonable to expect that $\mathcal{D}(s)$ contains interactions that decay as a power-law in distance, as inherited from $H(s)$. If so, then we expect to be able to prove a result analogous to Theorem 1, guaranteeing that the entanglement entropy $S_{V}(|\psi_{0}(s)\rangle)$ satisfies the dynamical area law $\left|dS_{V}(|\psi_{0}(s)\rangle)/ds\right|\le\mathcal{O}(|\partial V|)$ for $\alpha$ larger than a certain critical value. Upon integrating from $s=0$ to $s=1$, this would lead immediately to our Theorem 2 \cite{1_footnote}: 
\begin{thm}
(Area law for ground states) For $H(s)$ defined in Eq.\,\eqref{eq:Hs} with $\alpha\ge2D+2$, $|\psi_{0}(0)\rangle$ satisfying the area law implies that $|\psi_{0}(s)\rangle$ satisfies the area law for any $s\in[0,1]$. 
\end{thm}
Here the assumption that $|\psi_{0}(0)\rangle$ satisfies the area law may come from the scenario where $H(0)$ contains only short-range interactions. The proof of this area law is much more challenging than the proof of Theorem 1. To see the challenge, let us write $H(s)=\sum_{ij}h_{ij}(s)$ and $\mathcal{D}(s)=\sum_{ij}\mathcal{D}_{ij}(s)$, then \vspace{-1bp}
\begin{equation}
\mathcal{D}_{ij}(s)=-i\int_{-\infty}^{\infty}f(\Delta t)\tilde{h}_{ij}^{(s)}(t)dt,\label{eq:Dij}
\end{equation}
with $\tilde{h}_{ij}^{(s)}(t)\equiv e^{iH(s)t}\tilde{h}_{ij}e^{-iH(s)t}$ and $\tilde{h}_{ij}\equiv h_{ij}(1)-h_{ij}(0)$. Unlike $h_{ij}(s)$, which acts only on sites $i$ and $j$, in general $\mathcal{D}_{ij}(s)$ acts on the entire lattice. Thus we cannot directly apply Lemma 1 to constrain the growth of $S_{V}(|\psi(s)\rangle)$, as we did for Theorem 1. To overcome this challenge, we need to derive some locality structure of the interaction $\mathcal{D}_{ij}(s)$ despite the fact that it acts on the entire lattice. As mentioned above, our intuition is that $\mathcal{D}_{ij}(s)$ should be similar to $h_{ij}(s)$, in that it ``mostly'' acts on sites close to $i$ and $j$ while its interaction strength should still decay as $1/r_{ij}^{\alpha}$. In order to turn this intuition into a precise statement, we need to first look at the locality structure of $A(t)=e^{iHt}Ae^{-iHt}$ for $A$ acting on a set of sites $X$ and $H$ defined in Eq.\,\eqref{eq:H}.

Formally, we will define $A(t,R)=\int d\mu(U_{R})U_{R}A(t)U_{R}^{\dagger}$, with $U_{R}$ being a unitary operator acting on all sites with distance larger than or equal to $R$ from any site in $X$ and $\mu(U_{R})$ being the Haar measure for $U_{R}$. By this definition, $A(t,R)$ only acts on sites within a distance $R$ from $\partial X$. Let us first obtain some intuition in the $\alpha\rightarrow\infty$ limit, where $H$ is a nearest-neighbor Hamiltonian. It is reasonable to expect that $A(t,R)$ is a good approximation of $A(t)$ if we choose $R\gg t$, because it takes a time $t\propto R$ to ``spread'' the operator $A$ to sites a distance $R$ from its boundary. More precisely, one can apply the Lieb-Robinson bound \cite{lieb_finite_1972,bravyi_lieb-robinson_2006} in this case to obtain $\left\Vert A(t)-A(t,R)\right\Vert \le\left\Vert A\right\Vert \mathcal{O}(e^{4et-R})$. In fact, in the limit of $\alpha\rightarrow\infty$, Theorem 2 has already been proven in Ref.\,\cite{van_acoleyen_entanglement_2013}.

For a finite $\alpha$ the situation is much less clear. Using the direct generalization \cite{hastings_spectral_2006,nachtergaele_lieb-robinson_2010} of the Lieb-Robinson bound for the $1/r^{\alpha}$ Hamiltonian in Eq.\,\eqref{eq:H} leads to $\left\Vert A(t)-A(t,R)\right\Vert \le\left\Vert A\right\Vert |X|\mathcal{O}(e^{vt}/R^{\alpha-D})$, which only guarantees that $A(t)$ will be well approximated by $A(t,R)$ when $t\ll\log(R)$, thus requiring \emph{exponentially} larger $R$ to maintain the level of approximation in the $\alpha\rightarrow\infty$ case. As shown later, this requirement $t\ll\log(R)$ prohibits a proof of Theorem 2 using the strategy of Ref.\,\cite{van_acoleyen_entanglement_2013}. However, recent improvements to the long-range Lieb-Robinson bound \cite{foss-feig_nearly_2015} significantly improve the situation. The improved bound enables the following Lemma to be derived (see \cite{0_supp}), which together with additional techniques described below leads to a proof of Theorem 2. 
\begin{lem}
There exists a constant $v\!=\!\mathcal{O}(1)$ such that for $\alpha>2D$, $\gamma=\frac{D+1}{\alpha-2D}$, and $0<t<t_{R}\equiv(\frac{R}{6v})^{\frac{1}{1+\gamma}}$ \cite{2_footnote},\vspace{-10bp}
 
\begin{equation}
\left\Vert A(t)\!-\!A(t,R)\right\Vert \!\le\!\left\Vert A\right\Vert \!|X|[\mathcal{O}(e^{vt-\frac{R}{t^{\gamma}}})+\mathcal{O}(\frac{t^{\alpha(1+\gamma)}}{R^{\alpha-D}})].\label{eq:LRLR}
\end{equation}
\end{lem}
\noindent A crucial consequence of Lemma 2 is that we must only choose $R$ \emph{polynomially} large in $t$ in order to ensure that $A(t)$ is well approximated by $A(t,R)$. The quantity $t_{R}$ characterizes the edge of the ``light cone'', meaning that $\left\Vert A(t)-A(t,R)\right\Vert $ is only parametrically small in $R$ for $t<t_{R}$.

\begin{figure}
\includegraphics[width=0.5\textwidth,height=0.21\textheight]{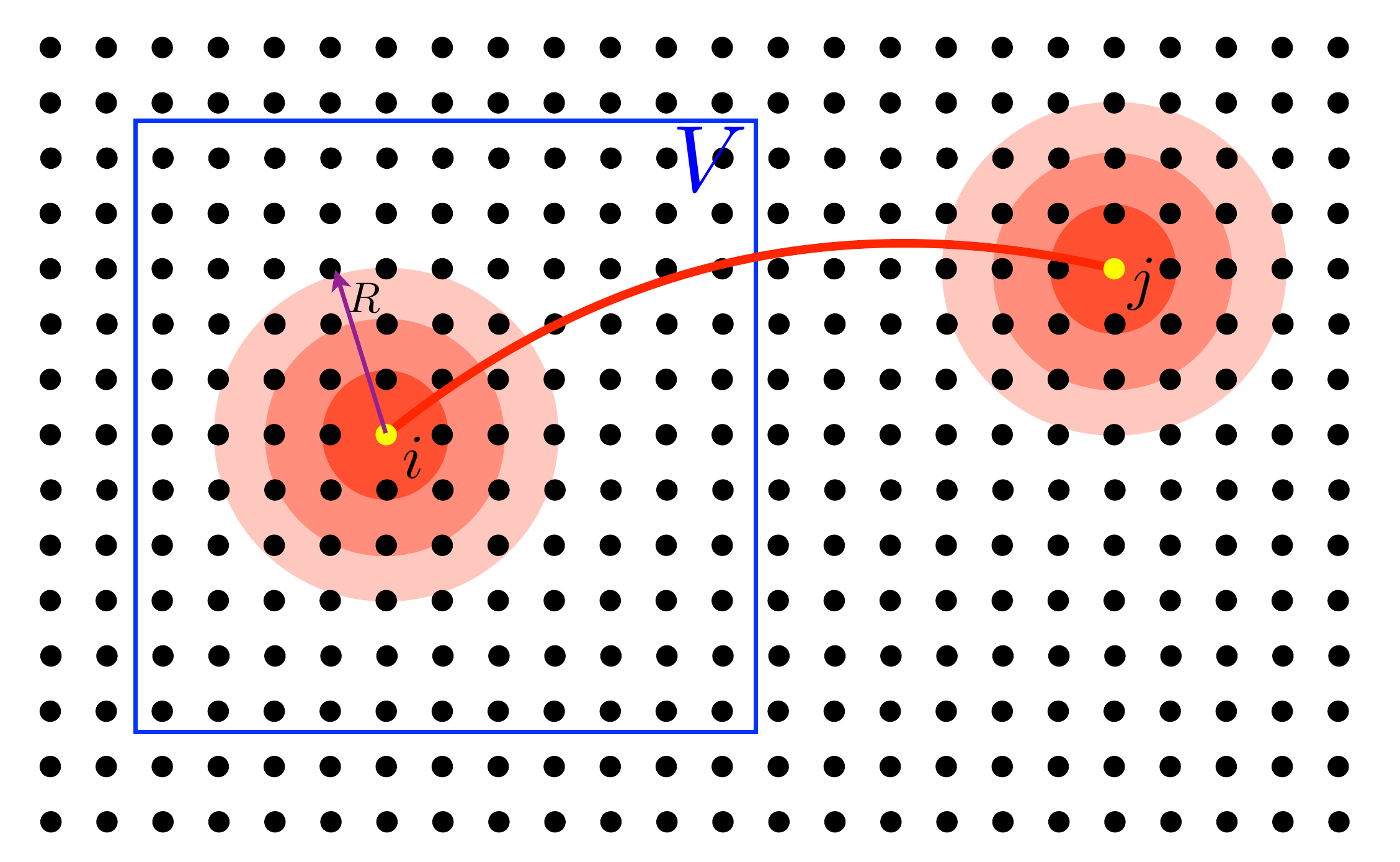} \caption{\label{fig:1}Illustration of the locality structure of $\mathcal{D}(s)$. Each $\mathcal{G}_{ij}(s,R)$ is an interaction between a ball of sites centered on $i$ with a radius $R$ and a ball of sites centered on $j$ with a radius $R$. The interaction strength $\lVert\mathcal{G}_{ij}(s,R)\rVert$ decays as $1/r_{ij}^{\alpha}$ and also as $1/R^{D-\alpha}$ for large $R$, represented by the fading color of the balls. For a given subregion $V$ with boundary $\partial V$ (blue square), the maximum rate of entanglement entropy change only involves interactions $\mathcal{G}_{ij}(s,R)$ that act on both sites in $V$ and sites outside $V$.\vspace{-5bp}
 }
\end{figure}

The locality structure of $\mathcal{D}_{ij}(s)$ can be understood with the help of Lemma 2 and the decomposition, 
\begin{equation}
\mathcal{D}_{ij}(s)\!=\!\sum_{R=1}^{\infty}\mathcal{G}_{ij}(s,R)\equiv-i\sum_{R=1}^{\infty}\!\int_{-\infty}^{\infty}\!f(\Delta t)g_{ij}^{(s)}(t,R)dt.\label{eq:Gij}
\end{equation}
Here, $g_{ij}^{(s)}(t,R)\equiv\tilde{h}_{ij}^{(s)}(t,R)-\tilde{h}_{ij}^{(s)}(t,R-1)$ and $\tilde{h}_{ij}^{(s)}(t,R)\equiv\int d\mu(U_{R})U_{R}\tilde{h}_{ij}^{(s)}(t)U_{R}^{\dagger}$; Eq.\,\eqref{eq:Gij} follows by bringing the summation inside the integral, and using $\tilde{h}_{ij}^{(s)}(t,\infty)\!=\!\tilde{h}_{ij}^{(s)}(t)$ and $\tilde{h}_{ij}^{(s)}(t,0)\!=\!0$ to collapse the summation to $\sum_{R=1}^{\infty}g_{ij}^{(s)}(t,R)=\tilde{h}_{ij}^{(s)}(t)$. We emphasize that $\mathcal{G}_{ij}(s,R)$ acts only on sites within a distance $R$ from $i$ or $j$ (Fig.\,\ref{fig:1}), and in this sense is ``local''. In order to bound how $\lVert\mathcal{G}_{ij}(s,R)\rVert$ decays with $R$ and $r_{ij}$, we must first derive a bound for $\lVert g_{ij}^{(s)}(t,R)\rVert$. It is useful to tackle the short-time and long-time behavior separately.  

\emph{Short-time behavior}: For $0<t<t_{R}$ we can apply Lemma 2. Using a triangle inequality $\lVert g_{ij}^{(s)}(t,R)\rVert\le\lVert\tilde{h}_{ij}^{(s)}(t)-\tilde{h}_{ij}^{(s)}(t,R)\rVert+\lVert\tilde{h}_{ij}^{(s)}(t)-\tilde{h}_{ij}^{(s)}(t,R-1)\rVert$  and the inequalities $vt-R/t^{\gamma}<-\mathcal{O}(R^{1/(1+\gamma)})$ and $\lVert\tilde{h}_{ij}\rVert\le2r_{ij}^{-\alpha}$, Lemma 2 gives $\lVert g_{ij}^{(s)}(t,R)\rVert\le[\mathcal{O}(e^{-\mathcal{O}(R^{\frac{1}{1+\gamma}})})+\mathcal{O}(\frac{t^{\alpha(1+\gamma)}}{R^{\alpha-D}})]r_{ij}^{-\alpha}.$

\emph{Long-time behavior}: When $t>t_{R}$, for reasons that will become clear soon it suffices to bound $\lVert g_{ij}^{(s)}(t,R)\rVert$ directly by $2\lVert\tilde{h}_{ij}\rVert\le4r_{ij}^{-\alpha}$, which follows because $\lVert A(t,R)\rVert=\lVert A\rVert$ for any $A$, t and $R$.

Performing the integration over $t$ in the definition of $\mathcal{G}_{ij}(s,R)$ {[}see Eq.\,\eqref{eq:Gij}{]}, we find \cite{3_foonote}\vspace{-5bp}
\[
\lVert\mathcal{G}_{ij}(s,R)\rVert\le\frac{\mathcal{O}(e^{-\mathcal{O}(R^{\frac{1}{1+\gamma}})})+\mathcal{O}(R^{D-\alpha})+\mathcal{O}(F[\mathcal{O}(t_{R})])}{r_{ij}^{\alpha}},
\]
where $F(x)=\int_{x}^{\infty}f(t)dt$ also decays sub-exponentially. Importantly, because Lemma 2 states that $t_{R}=\mathcal{O}(R^{1/(1+\gamma)})$, $\lVert\mathcal{G}_{ij}(s,R)\rVert$ is dominated by $\mathcal{O}(R^{D-\alpha})$ for large $R$. Note that the directly generalized Lieb-Robinson bound in Refs.\,\cite{hastings_spectral_2006,nachtergaele_lieb-robinson_2010} gives $t_{R}\sim\log(R)$; in this case, the term $\mathcal{O}(F[\mathcal{O}(t_{R})])$ above would not decay in $R$ for large $R$. 

To summarize what we have obtained so far, 
\begin{equation}
\mathcal{D}(s)\!=\!\sum_{ij}\sum_{R=1}^{\infty}\mathcal{G}_{ij}(s,R),\,\left\Vert \mathcal{G}_{ij}(s,R)\right\Vert \le\frac{\mathcal{O}(R^{D-\alpha})}{r_{ij}^{\alpha}}.\label{eq:DijR}
\end{equation}
Eq.\,\eqref{eq:DijR} reveals the locality structure hidden in $\mathcal{D}(s)$ (see Fig.\,\ref{fig:1} for an illustration); Theorem 2 can now be proved using Lemma 1 by summing over all $\lVert\mathcal{G}_{ij}(s,R)\rVert$ whose support overlap with $V$ and $\bar{V}$ simultaneously. Our summation strategy is to first sum over all $i$ and $j$ that contribute to $\left|dS_{V}(|\psi_{0}(s)\rangle)/ds\right|$ for a given $R$, and sum over $R$ next. The first step involves two scenarios: (1) For $i$ with $r_{i}\le R$ we need to sum $j$ over the entire lattice because $\mathcal{G}_{ij}(s,R)$ will always cross the boundary, leading to the summation $\sum_{i,r_{i}<R}\sum_{j}r_{ij}^{-\alpha}\sim R|\partial V|$ for $\alpha>D$. (2) For $i\in V$ and $r_{i}>R$, we will sum $j$ over sites with $r_{ij}>r_{i}-R$, corresponding to the summation $\sum_{i,r_{i}>R}\sum_{j,r_{ij}>r_{i}-R}r_{ij}^{-\alpha}\sim|\partial V|$ for $\alpha>D+1$. Therefore,\vspace{-10bp}

\begin{align}
\left|\frac{dS_{V}(|\psi_{0}(s)\rangle)}{ds}\right| & \le\sum_{R=1}^{\infty}R|\partial V|\mathcal{O}(R^{D-\alpha})R^{D},\label{eq:SIE-1}
\end{align}
where the final $R^{D}$ comes from the number of sites that $\mathcal{D}_{ij}(s,R)$ acts on. The summation converges for $\alpha>2D+2$, proving Theorem 2.

Note that the critical values of $\alpha$ in Theorems 1 and 2 differ by $D+1$, despite the fact that both $\lVert h_{ij}\rVert$ and $\lVert\mathcal{D}_{ij}(s)\rVert$ are bounded by $\mathcal{O}(r_{ij}^{-\alpha})$. This difference can be attributed to two differences between the locality structures of $\mathcal{D}(s)$ and $H$: (1) Each $\mathcal{G}_{ij}(s,R)$ acts on $\mathcal{O}(R^{D})$ sites while each $h_{ij}$ only acts on two sites. (2) There is an extra summation over the one-dimensional variable $R$ in $\mathcal{D}_{ij}(s)$.

\emph{Outlook}.— For the dynamical area law, an intriguing question is whether the area law can be extended to $\alpha<D+1$. Suppose the linear size of the subregion $V$ is $L$ and $|V|\propto L^{D}$, then from the proof of Theorem 1 one finds that $|dS_{V}/dt|\le\mathcal{O}(L^{2D-\alpha})$ for $\alpha\le D+1$. While this bound allows the area law to be violated, saturating it requires that each interaction $h_{ij}$ in Eq.\,\eqref{eq:H} provides the maximum (or a finite portion of the maximum) entanglement rate. Recently, a protocol using all $h_{ij}$ in Eq.\,\eqref{eq:H} was found for creating a single pair of entangled qubits separated by a distance $L$ in a $D$-dimensional lattice, and requires a time $t\propto L^{\alpha-D}$ \cite{eldredge_fast_2016} for $D<\alpha<D+1$ and a constant time for $\alpha<D$. If such a protocol can be generalized to apply in parallel for all the qubits in $V$, then $|dS_{V}/dt|=\mathcal{O}(L^{2D-\alpha})$ is achieved. However it seems plausible that the parallelization of this protocol may violate the monogamy of entanglement \cite{Coffman2000}. We leave the \emph{de facto} upper limit on the entanglement rate for Eq.\,\eqref{eq:H} as an open question.

Similarly, it remains unclear whether the critical value of $\alpha_{c}=2D+2$ is optimal in our ground-state area law. While the specific value of $\alpha_{c}$ may not have a fundamental importance so long as a finite $\alpha_{c}$ exists, for many experimental systems such as the $1/r^{6}$-interacting Rydberg atoms and $1/r^{3}$-interacting dipolar systems, knowing the smallest possible value of $\alpha_{c}$ can be crucial for deciding whether certain topological phases remain stable in the presence of long-range interactions \cite{yao_quantum_2015,yao_topological_2012,yao_realizing_2013}. We can, however, rule out the relevance of improving Lemma 2. As mentioned in the outlook of Ref.\,\cite{foss-feig_nearly_2015}, the long-range Lieb-Robinson bound obtained there, which is the basis of Lemma 2, is most likely not optimal. The best improvement of the long-range bound one could hope for is to demonstrate a linear light cone for $\alpha>D+1$ \cite{eldredge_fast_2016}. However, such a bound would not improve the value of $\alpha_{c}$ in Theorem 2, because the locality structure of $\mathcal{D}(s)$ {[}see Eq.\,\eqref{eq:DijR}{]} remains intact so long as a polynomial light cone is implied in Lemma 2. We also point out that the $1/R^{\alpha-D}$ decay in Lemma 2 cannot be improved further \cite{0_supp}.

Finally, Theorem 2 tells us the adiabatically connected ground states have similar entanglement properties. But do these ground states actually belong to the same quantum phase? The answer is known to be yes for short-range interacting systems \cite{chen_local_2010-1}, but is not yet clear if interactions are long-ranged. In addition, will the proved stability of the area law imply the stability of topological orders \cite{bravyi_lieb-robinson_2006}? We believe that our results will help obtain a more general understanding of the emergent notion of locality that underpins a wide range of many-body physics in long-range interacting systems. 
\begin{acknowledgments}
We thank M.\,Hastings, G. Zhu, and R. Lundgren for helpful discussions. This work was supported by the AFOSR, NSF QIS, ARL CDQI, ARO MURI, ARO, NSF PFC at the JQI. Z.-X. Gong thanks the PFC seed grant at JQI for support. 
\end{acknowledgments}

\bibliographystyle{apsrev4-1}
\bibliography{sub}

\begin{thebibliography}{63}
\bibitem [{2_f()}]{2_footnote}%
  \BibitemOpen
  \href@noop {} {}\bibinfo {note} {Strictly speaking, one should replace $t$ by
  $\lceil t\rceil$ (the smallest integer larger than $t$) in Lemma 2. But this
  nuance does not affect our main results.}\BibitemShut {Stop}%
\bibitem [{3_f()}]{3_foonote}%
  \BibitemOpen
  \href@noop {} {}\bibinfo {note} {Note that the integration just needs to be
  performed for $t>0$, because for $t<0$ we can simply flip $H(s)$ to $-H(s)$
  and $\lVert g_{ij}^{(s)}(-t,R)\rVert$ is subject to the same bounds as for
  $t>0$.}\BibitemShut {Stop}%
\end{thebibliography}

\onecolumngrid
\newpage

\begin{center}
\Large\textbf{Supplemental Material for ``Entanglement area laws for long-range interacting systems''}
\end{center}

\setcounter{equation}{0}
\renewcommand{\theequation}{S\arabic{equation}}

In this supplemental material, we prove Lemma 2 in the main text by generalizing the following result of Ref.\,\cite{foss-feig_nearly_2015}:
\begin{quote}
\emph{For any operators $A$ and $B$ acting on sites $i$ and $j$ respectively, there exists constants $v=\mathcal{O}(1)$ such that for $\alpha>2D$, $\gamma=\frac{D+1}{\alpha-2D}$, and $0<t<t_{R}\equiv(\frac{R}{6v})^{\frac{1}{1+\gamma}}$,} 
\begin{equation}
\left\Vert [A(t),B\right\Vert \le\left\Vert A\right\Vert \left\Vert B\right\Vert \left[\mathcal{O}(e^{vt-r_{ij}/t^{\gamma}})+\mathcal{O}(\frac{t^{\alpha(1+\gamma)}}{r_{ij}})\right].\label{eq:Mike's}
\end{equation}
\end{quote}
\noindent In order to prove Lemma 2 in the main text, the above Lieb-Robinson-type bound must be extended to allow for operators $A$ and $B$ that are supported on an arbitrary number of sites. To proceed we will need to recall how Eq.\,\eqref{eq:Mike's} is derived. Let us first quote Eq.\,(11-12) in Ref.\,\cite{foss-feig_nearly_2015}, 
\begin{align}
\left\Vert [A(t),B]\right\Vert  & \leq\sum_{\ell}\left[\lVert[\mathcal{A}^{\ell}(t),B]\rVert+4c\sum_{k=1}^{\infty}\frac{t^{a}}{a!}\mathcal{J}_{a}(i,j)\right],\label{eq:series}\\
\mathcal{J}_{a}(i,j) & =4^{n}\sum_{\ell,\xi_{1},\dots,\xi_{a}}e^{-\ell}D_{{\rm i}}(\xi_{1})\lVert\mathcal{W}_{\xi_{1}}\rVert D(\xi_{1},\xi_{2})\lVert\mathcal{W}_{\xi_{2}}\rVert\times\dots\dots\times\lVert\mathcal{W}_{\xi_{a-1}}\rVert D(\xi_{a-1},\xi_{a})\lVert\mathcal{W}_{\xi_{a}}\rVert D_{{\rm f}}(\xi_{a}).\label{eq:Ja}
\end{align}
We will not introduce all of the notation used in the above two equations, as it can be found Ref.\,\cite{foss-feig_nearly_2015}, but we will explain the aspects of the notation that are relevant for our purposes. The operator $\mathcal{A}^{\ell}(t)$ is exclusively supported on the set $\mathcal{B}_{\ell}(i)$, which is the set of sites with distance $\le(vt+\ell)\chi$ from site $i$, with $v=\mathcal{O}(1)$ the so-called Lieb-Robinson velocity. Importantly, $\left\Vert \mathcal{A}^{l}(t)\right\Vert \le\left\Vert A\right\Vert \mathcal{O}(e^{-l})$ (see Eq.\,(S8) in Ref.\,\cite{foss-feig_nearly_2015}). The collective index $\xi_{k}=(x_{k},y_{k},m_{k},n_{k})$ specifies the support of the operator $\mathcal{W}_{\xi_{k}}$, which is an interaction connecting sites in $\mathcal{B}_{m_{k}}(x_{k})$ to sites in $\mathcal{B}_{n_{k}}(y_{k})$ (and is supported on the unions of these two sets). The quantity $D_{i}(\xi_{1})=1$ when $\mathcal{B}_{\ell}(i)\cap\mathcal{B}_{m_{1}}(x_{1})\neq\varnothing$ and vanishes otherwise, while the quantity $D_{{\rm f}}(\xi_{a})$ is unity when $j\in\mathcal{B}_{n_{a}}(y_{a})$ and vanishes otherwise. Similarly, $D(\xi_{k-1},\xi_{k})=1$ when $\mathcal{B}_{n_{k-1}}(y_{k-1})\cap\mathcal{B}_{m_{k}}(x_{k})\neq\varnothing$, and vanishes otherwise. Intuitively, these quantities ensure that $\mathcal{J}_{a}(i,j)$ connects operators $A$ to $B$ and contributes to the commutator $[A(t),B]$.

Now we assume that $A$ and $B$ act on two arbitrary sets $X$ and $Y$ respectively, which results in a number of changes. First, $\mathcal{A}^{\ell}(t)$ will instead act on $\mathcal{B}_{\ell}(X)$, which is defined as the set of sites with distance $\le(vt+\ell)\chi$ from any site in $X$. But $\left\Vert \mathcal{A}^{l}(t)\right\Vert \le\left\Vert A\right\Vert \mathcal{O}(e^{-l})$ holds independent of the size of $X$, because it is obtained using a finite-range Lieb-Robinson bound \cite{gong_persistence_2014}. Second, $D_{i}(\xi_{1})$ in Eq.\,\eqref{eq:Ja} should now be replaced by $D_{X}(\xi_{1})$, which restricts the summation over $\xi_{1}$ to the cases where $\mathcal{B}_{\ell}(X)\cap\mathcal{B}_{m_{1}}(x_{1})\neq\varnothing$. Finally, $D_{j}(\xi_{a})$ should be replaced by $D_{Y}(\xi_{a}$) such that the summation over $\xi_{a}$ is restricted such that $\mathcal{B}_{\ell}(Y)\cap\mathcal{B}_{n_{a}}(y_{a})\neq\varnothing$.

Next, we observe that the summation over $\xi_{1}$ with the constraint $\mathcal{B}_{\ell}(X)\cap\mathcal{B}_{m_{1}}(x_{1})\neq\varnothing$ is upper bounded by the summation over $\xi_{1}$ with the constraint $\mathcal{B}_{\ell}(i)\cap\mathcal{B}_{m_{1}}(x_{1})\neq\varnothing$ \emph{plus} an extra summation over all $i\in X$. This is true because for $\mathcal{B}_{m_{1}}(x_{1})$ to overlap with $\mathcal{B}_{\ell}(X)$, it has to overlap with $\mathcal{B}_{\ell}(i)$ for some $i\in X$. In summing over all $i\in X$, we may count the same $\mathcal{B}_{m_{1}}(x_{1})$ that overlaps multiple times, but since the summand in Eq.\,\eqref{eq:Ja} is always non-negative we nevertheless obtain an upper bound. A similar treatment will be applied to the summation over $\xi_{a}$ as well.

As a result, we will replace $\mathcal{J}_{a}(i,j)$ in Eq.\,\eqref{eq:series} by $\sum_{i\in X,j\in Y}\mathcal{J}_{a}(i,j)$. The summation $\sum_{\ell}\lVert[\mathcal{A}^{\ell}(t),B]\rVert$ in Eq.\,\eqref{eq:series} is now restricted only to $\ell$s satisfying $(vt+\ell)\chi\ge r_{XY}$, These two changes together lead to a modified version of Eq.\,\eqref{eq:Mike's}, which reads 
\begin{equation}
\left\Vert [A(t),B]\right\Vert \leq\left\Vert A\right\Vert \left\Vert B\right\Vert \left[\mathcal{O}(e^{vt-r_{XY}/t^{\gamma}})+\sum_{i\in X,j\in Y}\mathcal{O}(\frac{t^{\alpha(1+\gamma)}}{r_{ij}^{\alpha}})\right].\label{eq:final_bound}
\end{equation}
Finally, we set $B=U_{R}$, which is an arbitrary unitary acting only on sites with distance larger than or equal to $R$ from all sites in $X$, and obtain (whenever $\alpha>D$) 
\begin{equation}
\left\Vert [A(t),U_{R}]\right\Vert \le\left\Vert A\right\Vert |X|[\mathcal{O}(e^{vt-R/t^{\gamma}})+\mathcal{O}(\frac{t^{\alpha(1+\gamma)}}{R^{\alpha-D}})].\label{eq:LRLR}
\end{equation}
Using $\left\Vert A(t)-A(t,R)\right\Vert \le\int d\mu(U_{R})\left\Vert [A(t),U_{R}]\right\Vert $, we have proven Lemma 2 in the main text.

As mentioned towards the end of the main text, it can be shown that Eq.\,\eqref{eq:final_bound} is the optimal generalization of Eq.\,\eqref{eq:Mike's}. To see this, consider a long-range Ising model $H=\sum_{ij}J_{ij}\sigma_{i}^{z}\sigma_{j}^{z}$ with $J_{ij}=r_{ij}^{-\alpha}$ if $i\in X$ and $j\in Y$, and $J_{ij}=0$ otherwise. We define 
\begin{equation}
|\psi_{0}\rangle=\left[\frac{1}{\sqrt{2}}\left(\bigotimes_{i\in X\cup Y}|\sigma_{i}^{z}=1\rangle+\bigotimes_{i\in X\cup Y}|\sigma_{i}^{z}=-1\rangle\right)\right]\bigotimes_{i\notin X\cup Y}|\sigma_{i}^{z}=1\rangle,
\end{equation}
which is a GHZ state for all sites in $X$ and $Y$. Now let us choose $A=\prod_{i\in X}\sigma_{i}^{+}$ and $B=\sum_{j\in Y}\sigma_{j}^{+}$. It is not hard to find that

\begin{align}
\left\langle \psi_{0}\right|\prod_{i\in X}\sigma_{i}^{+}(t)\prod_{j\in Y}\sigma_{j}^{+}\left|\psi_{0}\right\rangle  & =\frac{1}{2}e^{2i\sum_{i\in X,j\in Y}J_{ij}t},\\
\left\langle \psi_{0}\right|\prod_{j\in Y}\sigma_{j}^{+}\prod_{i\in X}\sigma_{i}^{+}(t)\left|\psi_{0}\right\rangle  & =\frac{1}{2}e^{-2i\sum_{i\in X,j\in Y}J_{ij}t}.
\end{align}
As a result, for $t<\pi/(4\sum_{i\in X,j\in Y}J_{ij})$, we have 
\begin{equation}
\left\Vert [A(t),B]\right\Vert \ge\left|\langle\psi_{0}|[A,B]|\psi_{0}\rangle\right|=\sin(2t\sum_{i\in X,j\in Y}J_{ij})>\frac{4}{\pi}t\sum_{i\in X,j\in Y}r_{ij}^{-\alpha}.
\end{equation}
Although the time dependence is different from Eq.\,\eqref{eq:final_bound}, the distance dependence (for large $r_{ij}$) and the summations over $i\in X$ and $j\in Y$ match those in Eq.\,\eqref{eq:final_bound}. Thus we claim that Eq.\,\eqref{eq:final_bound} is the optimal generalization of Eq.\,\eqref{eq:Mike's}.

\end{document}